\documentclass[%
  reprint,
  a4paper,
  aps,
  pra,
  amsmath,amssymb,
]{revtex4-2}

\usepackage[
  a4paper,
  top=2cm,
  bottom=2cm,
  left=2cm,
  right=2cm,
]{geometry}

\usepackage{graphicx}
\usepackage{dcolumn}
\usepackage{bm}
\usepackage{xcolor}
\usepackage{xurl}
\usepackage{subfiles}

\usepackage{hyperref}
\hypersetup{
  colorlinks = true,
  citecolor  = blue,
  linkcolor  = blue,
  urlcolor   = blue,
  pdftitle   = {Towards Hadronisation on GPUs},
  pdfauthor  = {Siddharth Sule},
}

\begin{document}

\title{Towards Hadronisation on GPUs}
\thanks{Release 2.2 of GAPS, a GPU-Amplified Parton Shower.}

\author{Siddharth Sule}
\email{siddharth.sule@manchester.ac.uk}
\affiliation{%
Department of Physics and Astronomy,\\
The University of Manchester, M13 9PL, United Kingdom
}%

\begin{abstract}
Complete event generation on hardware accelerators such as GPUs requires vectorising the algorithms of components such as parton showers and hadronisation. To this end, I implemented a simplified cluster hadronisation and hadronic decay model in the GPU parton shower program GAPS, with one thread per event and each stage of the algorithm as a separate kernel. I compared simulation results for $e^+e^- \to \text{hadrons}$ at the $Z$ pole with the Herwig event generator and with ALEPH and L3 data, and measured execution times on an NVIDIA V100 GPU. The results agree well with Herwig, and GAPS generates one million hadron-level events (NLO hard process, shower, hadronisation and decays) in about four seconds on this GPU.
\end{abstract}

\maketitle

\section{\label{sec:intro}Introduction}

In recent years, particle physics simulation tools have seen substantial progress in hardware acceleration \cite{Bothmann:2021nch, Bothmann:2023siu, Borowka:2018goh, Heinrich:2023til, Cruz-Martinez:2025kwa, Carrazza:2020qwu, Rossi:2020sbh, DeLaurentis:2026bhl}, notably event generators \cite{Bothmann:2023gew, Bothmann:2026yzp, Hagebock:2025jyk, Heimel:2026hgp, Valassi:2026pgd}.
For parton showers, the GAPS program is a proof of principle for initial- and final-state emissions, as well as NLO matching on NVIDIA GPUs \cite{Seymour:2024fmq, Seymour:2025fpu}.
While the key challenge in this field is computing NLO and NNLO events, parallelising non-perturbative components of event generators, such as hadronisation and multi-parton interactions, also brings hardware-accelerated event generation closer to production use.
In this note, I present a starting point towards this goal by porting cluster hadronisation and hadronic decay algorithms to GPUs, enabling GAPS to produce $e^+e^- \to \text{hadrons}$ events.
In my implementation, I followed the procedure used throughout the GAPS project: I did not change the algorithm for the GPU, but wrote each step as a CUDA kernel so many events can undergo hadronisation simultaneously.
I then benchmarked the physics against Herwig and LEP data and studied execution-time scaling with the number of events simulated in parallel.

\section{\label{sec:model}The Hadronisation Algorithm}

My implementation of the cluster hadronisation algorithm \cite{Webber:1983if} is based on the Herwig event generator \cite[ch. 7]{Bellm:2025pcw}.
With the simplifications described below, it is similar to the original Herwig++ algorithm \cite[ch. 7]{Bahr:2008pv}, and it does not include colour reconnection.
Each step of the algorithm, and how it differs from Herwig, is summarised below:

\textbf{Constituent Reshuffling:} In the setup used here, the hard process and parton shower in GAPS and Herwig use massless partons; hence, the final-state momenta are reshuffled onto the partons' constituent masses.
Gluons are assigned a fixed constituent mass $m_g$, although newer models draw the mass from a distribution \cite{Hoang:2018zrp, Hoang:2024nqi}.
GAPS reshuffles the momenta in the same way as Herwig, in the event rest frame, by solving
\begin{equation}
\sum_i |\mathbf{p}_i| = \sum_i \sqrt{K^2\mathbf{p}_i^2+m_i^2} \, ,
\end{equation}
for the reshuffling factor $K$, where $\mathbf{p}_i$ are the three-momenta of the massless partons and $m_i$ their constituent masses.
The momenta are then adjusted following $E_i \to \sqrt{K^2\mathbf{p}_i^2 + m_i^2}$ and $\mathbf{p}_i \to K\mathbf{p}_i$.

\textbf{Forced Gluon Splitting:} Each gluon is forced to split into a $d\bar{d}$, $u\bar{u}$ or $s\bar{s}$ pair, chosen with the flavour weights $w_q$ ($q \in \{d, u, s\}$), such that the final state contains only quarks and antiquarks.
As in Herwig, the splitting is isotropic in the gluon rest frame, with the momenta of the massive quarks given by the two-body momentum $p^*$, built from the K\"all\'en function $\lambda$ \cite[ch. 49]{ParticleDataGroup:2026mpi},
\begin{equation}
\label{eq:pstar}
\begin{gathered}
p^*(M, m_{1}, m_{2}) = \frac{\sqrt{\lambda\!\left(M^2, m_1^2, m_2^2\right)}}{2M} \, , \\
\lambda(x,y,z) = x^2 + y^2 + z^2 - 2xy - 2yz - 2zx  \, .
\end{gathered}
\end{equation}

\textbf{Cluster Formation:} Clusters are formed by pairing colour-connected quarks and antiquarks.

\textbf{Cluster Fission:} A cluster undergoes fission when it is too heavy, a condition defined as
\begin{equation}
M^{\texttt{Clpow}} \geq {\texttt{Clmax}}^{\texttt{Clpow}} + (m_1 + m_2)^{\texttt{Clpow}} \, ,
\end{equation}
where $M$ is the cluster mass, $m_1$ and $m_2$ are the masses of its constituents, and \texttt{Clmax} and \texttt{Clpow} are tunable parameters.
Following Herwig, I set the parameters separately for light, charm, and bottom clusters, since charm and bottom quark masses contribute significantly to the second term on the right-hand side.
In Herwig, this separation was introduced to improve the modelling of $b$-hadrons \cite{Corcella:2000bw}.
A $d\bar{d}$, $u\bar{u}$ or $s\bar{s}$ pair is drawn from the vacuum with the flavour weights $w_q$, and the cluster $(q_1\bar{q}_2)$ splits into two clusters $(q_1\bar{q})$ and $(q\bar{q}_2)$, emitted along the axis of $q_1$ in the cluster rest frame.
The masses of the two clusters are
\begin{equation}
\begin{aligned}
M_1 &= m_1 + (M-m_1-m_{q})\,\mathcal{R}_1^{1/\texttt{PSplit}} \, ,\\
M_2 &= m_2 + (M-m_2-m_{q})\,\mathcal{R}_2^{1/\texttt{PSplit}} \, ,
\end{aligned}
\end{equation}
where $m_q$ is the mass of the popped quark, $\mathcal{R}_1$ and $\mathcal{R}_2$ are uniform random numbers, and \texttt{PSplit} is a tunable parameter.
The flavour and the two random numbers are redrawn until $M_i \geq m_i + m_q$ and $M_1 + M_2 \leq M$.
The current default fission formulae in Herwig are more sophisticated \cite{Masouminia:2023zhb}; GAPS follows the older algorithm described above.

\textbf{Cluster Decay and Fallback:} Once no cluster requires fission, the clusters decay into hadrons, many of which are unstable.
Unlike cluster fission, cluster decay can produce baryons. 
A decay is chosen to be baryonic (cluster $\to$ baryon pair) with probability $w_{\text{diquark}}/(1 + w_{\text{diquark}})$, for the diquark weight $w_{\text{diquark}}$, and mesonic (cluster $\to$ meson pair) otherwise.
If a baryonic decay finds no kinematically allowed baryon pair, it falls back to a mesonic one.
A mesonic decay pops a $d\bar{d}$, $u\bar{u}$ or $s\bar{s}$ pair, and a baryonic decay pops a diquark--antidiquark pair, the diquark $qq'$ made of two of $u$, $d$ and $s$.
For every pair that can be popped, every pair of hadrons $(q_1\bar{q})(q\bar{q}_2)$ or $(q_1 qq')(\bar{q}\bar{q}'\,\bar{q}_2)$ stored in the particle database (explained in the next step) is assigned the weight
\begin{equation}
\begin{aligned}
W_{h_1 h_2} &= w_{\text{pop}} \, w_{h_1} \, w_{h_2} \, p^*(M, m_{h_1}, m_{h_2}) \, ,
\\
w_h &= (2J_h + 1)\,\eta_h \, , \\
w_{\text{pop}} &=
\begin{cases}
w_q & \text{mesonic} \, , \\
w_{\text{diquark}}\,w_q\,w_{q'} & \text{baryonic} \, ,
\end{cases}
\end{aligned}
\end{equation}
where $J_h$ is the spin of hadron $h$, and $\eta_h$ is the weight of the popped flavour in the hadron's flavour wavefunction.
This weight is 1 for a hadron of definite flavour content, ${1}/{2}$ for $u\bar{u}$ and $d\bar{d}$ in the neutral isovectors (such as $\pi^0$), and a mixing-angle factor for the isoscalars (such as $\eta$, $\eta'$); $ddd$, $uuu$, $sss$-like baryons carry an additional factor of $3/2$, as the Pauli exclusion principle allows only their $J = 3/2$ state.
Herwig's formula contains tunable suppression factors, which are omitted here.
Then, one pair is selected with probability proportional to its weight.
In terms of kinematics, GAPS decays clusters isotropically in their rest frame, an option also available in Herwig.
If a cluster is lighter than the lightest hadron pair it can produce, it becomes the lightest hadron its own constituents form, which is always a meson.
The cluster must then carry that meson's mass, so momentum is reshuffled: while Herwig exchanges momentum with the nearest neighbour in spacetime, GAPS rescales the momenta of the whole event, as in constituent reshuffling, before any cluster decays.

\textbf{Hadronic Decays:} Each unstable hadron, one with $c\tau$ below a threshold of $100\,\text{mm}$ by default, decays through one of its modes, chosen with probability proportional to its branching fraction, and its children are distributed flat over the $1 \to N$ phase space \cite{James:1968gu}.
For this, I generated the particle and decay databases using the Python packages \texttt{pdg}, \texttt{particle} and \texttt{decaylanguage} \cite{pdgapi, Rodrigues:2020syo, particle, decaylanguage}, which contain the masses and lifetimes that come from the Particle Data Group \cite{ParticleDataGroup:2026mpi}, and the decay modes and branching fractions from the EvtGen decay tables \cite{Lange:2001uf, Abudinen:2024wpc} from LHCb \cite{Clemencic:2011zza} and Belle~II \cite{Kuhr:2018lps}.
I skipped the partonic modes, whose children include quarks, gluons or diquarks, since their products would need hadronising in turn, which is beyond the scope of this release.
I removed every short-lived hadron with no usable decay mode, together with every mode decaying into one, until none remained, leaving the pseudoscalar, vector and tensor mesons, many of their orbital and radial excitations, and the octet and decuplet baryons, along with the few other states their decays produce.
In Herwig, hadron decays follow detailed models of the decay amplitudes and widths, and cover a much wider range of hadrons and decay modes \cite[ch. 9]{Bellm:2025pcw}.

\section{\label{sec:storage}Multiplicity and GPU Memory}

Incorporating hadronisation and hadronic decays significantly increases event multiplicity.
GAPS allocates each event record as a fixed array of $N_{\max}$ particle objects, which are filled as particles are generated in the hard process, shower, and hadronisation.
With the GAPS 2.1 default of $N_{\max} = 70$, $10\%$ of events overflowed this array.
Although an incomplete correction, increasing to $N_{\max} = 100$ reduced this to $\sim0.2\%$, and this is the new default (users can raise it further in \texttt{base.cuh}. However, a higher value reduces the number of events that can be simulated in parallel on a GPU such as the one used in Section~\ref{sec:times}).
An event that overflows is not processed further and is excluded from the histograms, but its weight still counts toward their normalisation; this slightly biases the high-multiplicity tail, which IRC-unsafe observables such as the charged multiplicity feel most, and I will address this in a future release.

A larger $N_{\max}$ also makes each event record larger, causing memory issues with the event-record partitioning algorithm introduced in~\cite{Seymour:2025fpu}, which reorders the event records on the device.
I therefore replaced it with a list of indices of `active' events, i.e.\ events that have not finished showering.
The shower kernels run over this list rather than the full event array, and after each shower cycle the list is filtered to remove finished events, so each cycle covers the same events as the original algorithm.
Here, though, only the integer indices move, so the event records stay in place, and no temporary copies are needed.
As in GAPS 2.1, this filtering stops once fewer than 25,000 events remain, as below this the overhead outweighs the speedup.

\begin{figure*}[!t]
\centering
\includegraphics[width=0.24\linewidth]{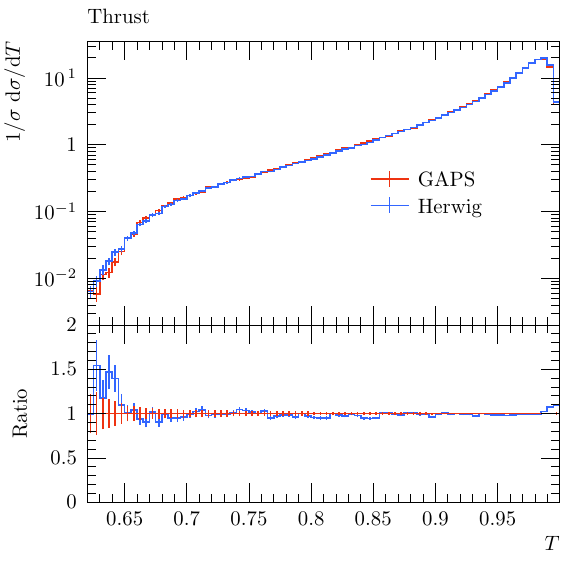}
\includegraphics[width=0.24\linewidth]{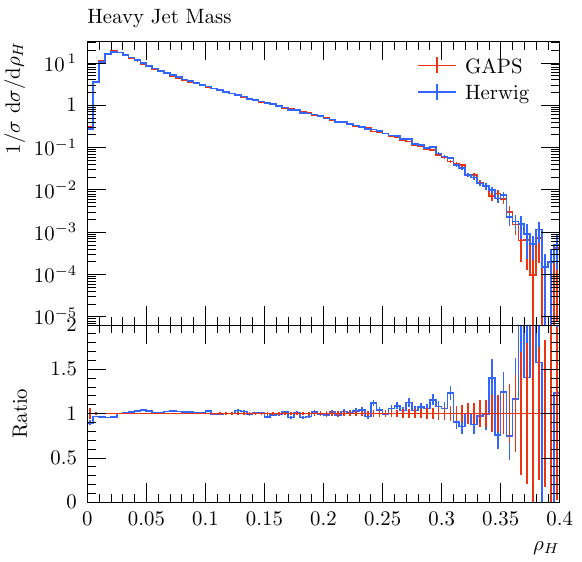}
\includegraphics[width=0.24\linewidth]{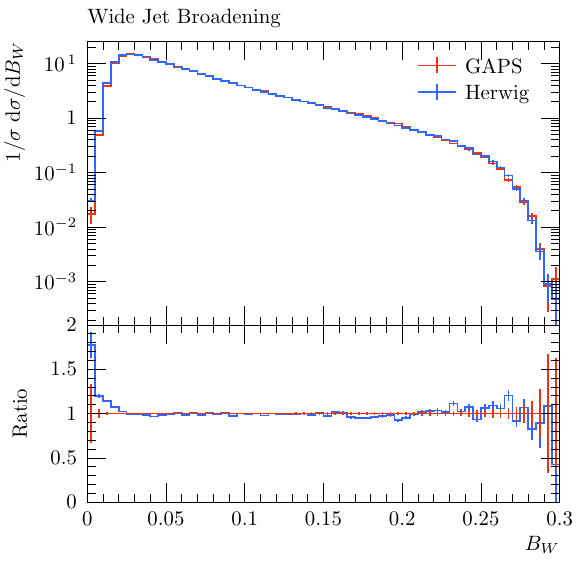}
\includegraphics[width=0.24\linewidth]{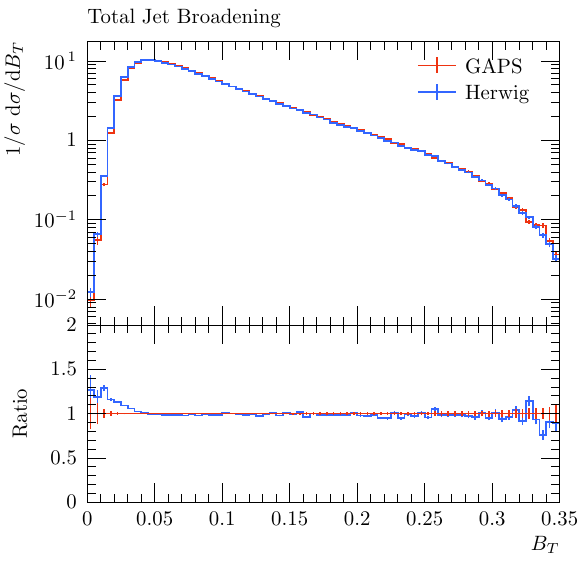}
\includegraphics[width=0.24\linewidth]{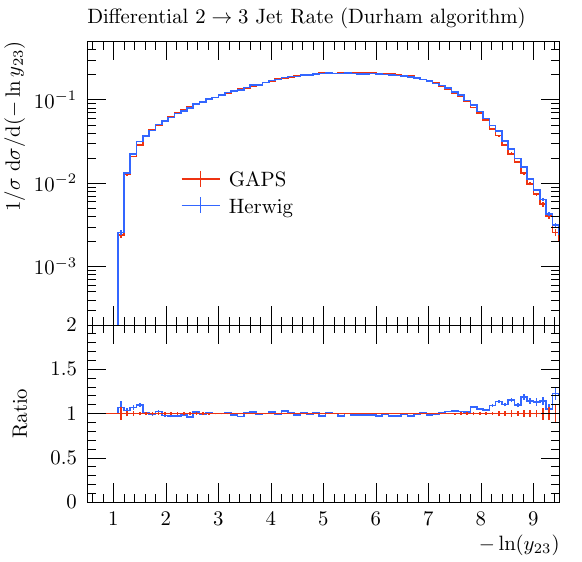}
\includegraphics[width=0.24\linewidth]{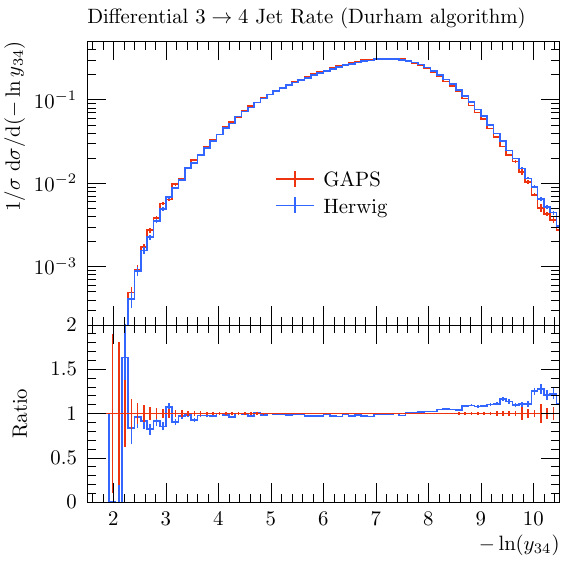}
\includegraphics[width=0.24\linewidth]{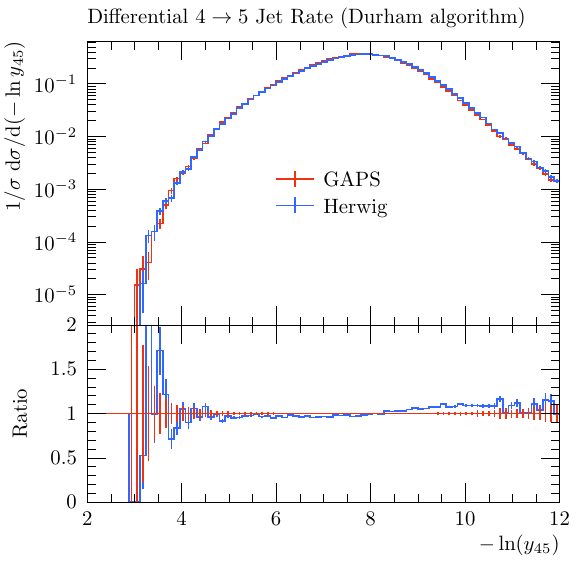}
\includegraphics[width=0.24\linewidth]{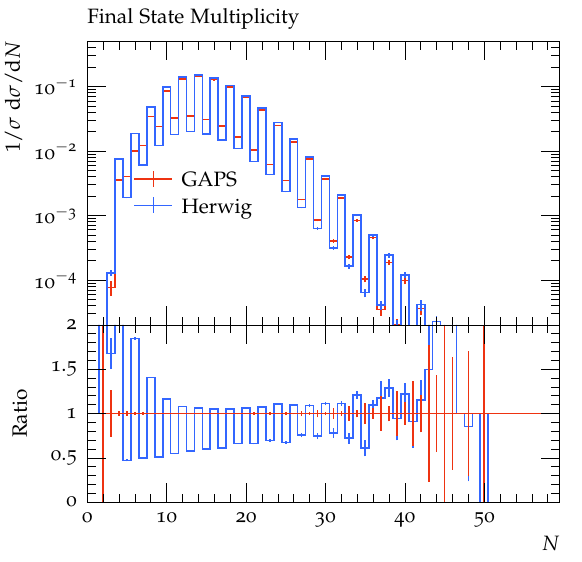}

\vspace{0.3em}
(a) NLO + shower + hadronisation, decays off.
\vspace{0.3em}

\includegraphics[width=0.24\linewidth]{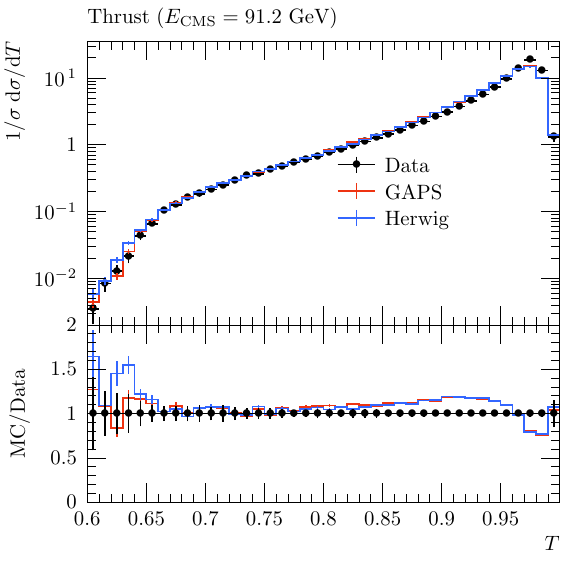}
\includegraphics[width=0.24\linewidth]{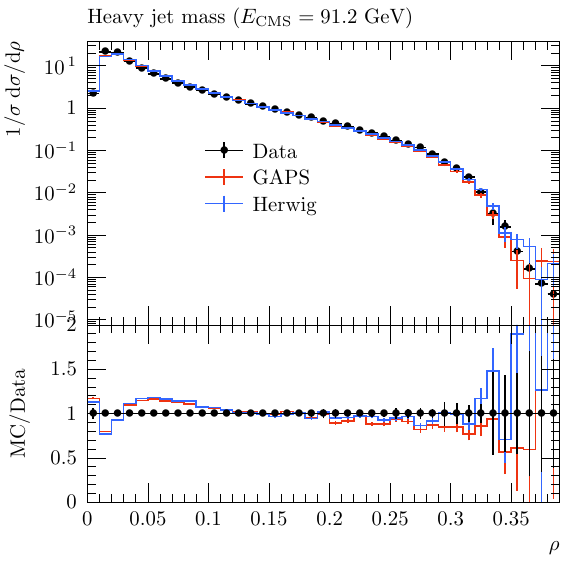}
\includegraphics[width=0.24\linewidth]{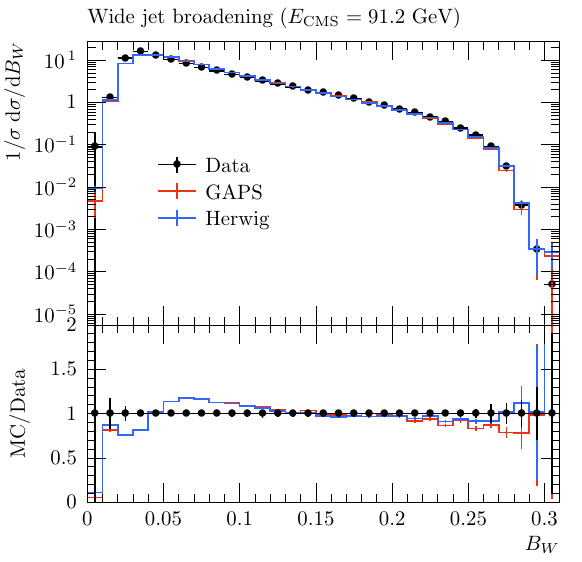}
\includegraphics[width=0.24\linewidth]{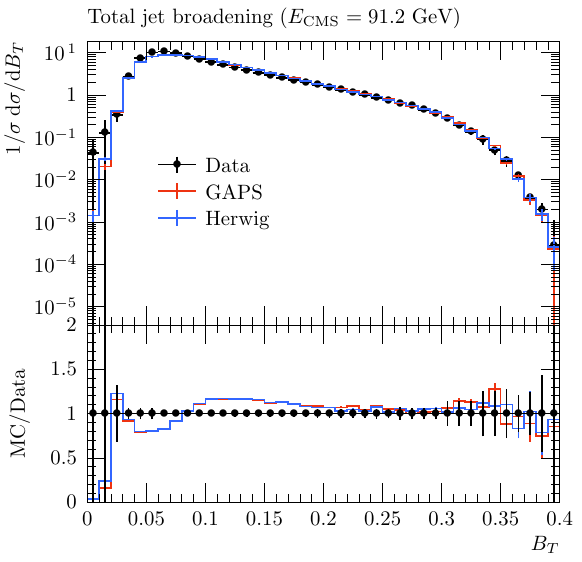}
\includegraphics[width=0.24\linewidth]{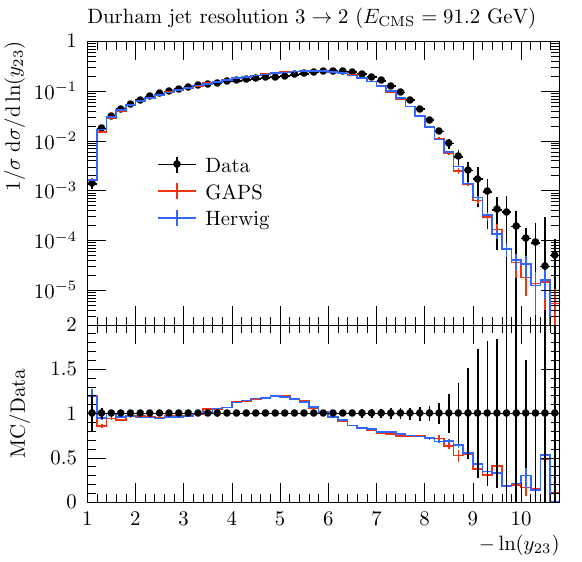}
\includegraphics[width=0.24\linewidth]{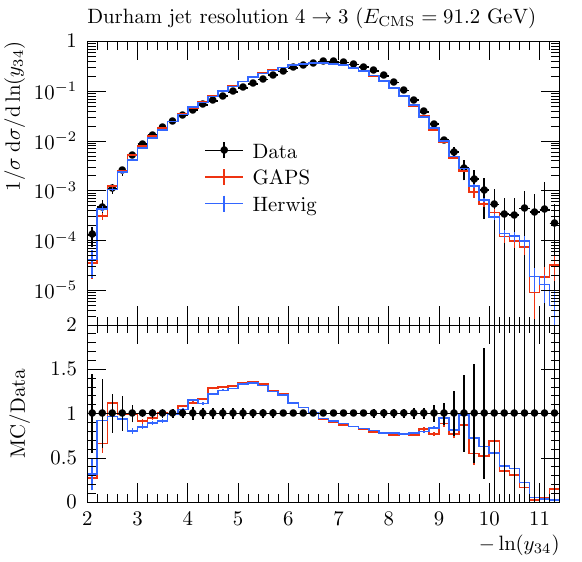}
\includegraphics[width=0.24\linewidth]{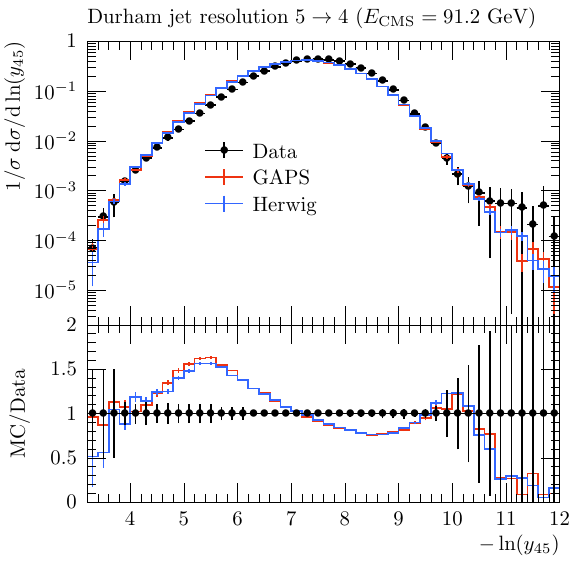}
\includegraphics[width=0.24\linewidth]{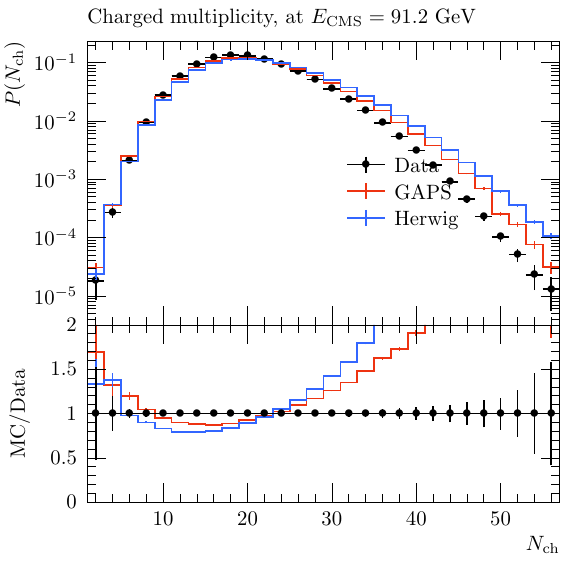}

\vspace{0.3em}
(b) NLO + shower + hadronisation + decays.
\vspace{0.3em}

\caption{%
Event shapes, jet rates and multiplicities, simulated with GAPS and Herwig.
(a) Without hadronic decays, GAPS and Herwig agree closely for all event shapes and jet rates, except in the two-jet limit, where GAPS lies $15\%$ below Herwig.
The multiplicity spikes at even $N$: gluon splitting, cluster fission, and cluster decay are all $1 \to 2$ processes, so an event contains an even number of hadrons unless a cluster undergoes single-hadron fallback.
GAPS produces more odd-multiplicity events than Herwig, suggesting more clusters become single hadrons.
(b) With hadronic decays, compared with ALEPH and L3 data; ratios are taken to the data.
These are not the best results achievable with Herwig, as colour reconnection and the new fission formulae are turned off.
}
\label{fig:hadron-level}
\end{figure*}

\begin{figure*}[!t]
\centering
\includegraphics[width=\linewidth]{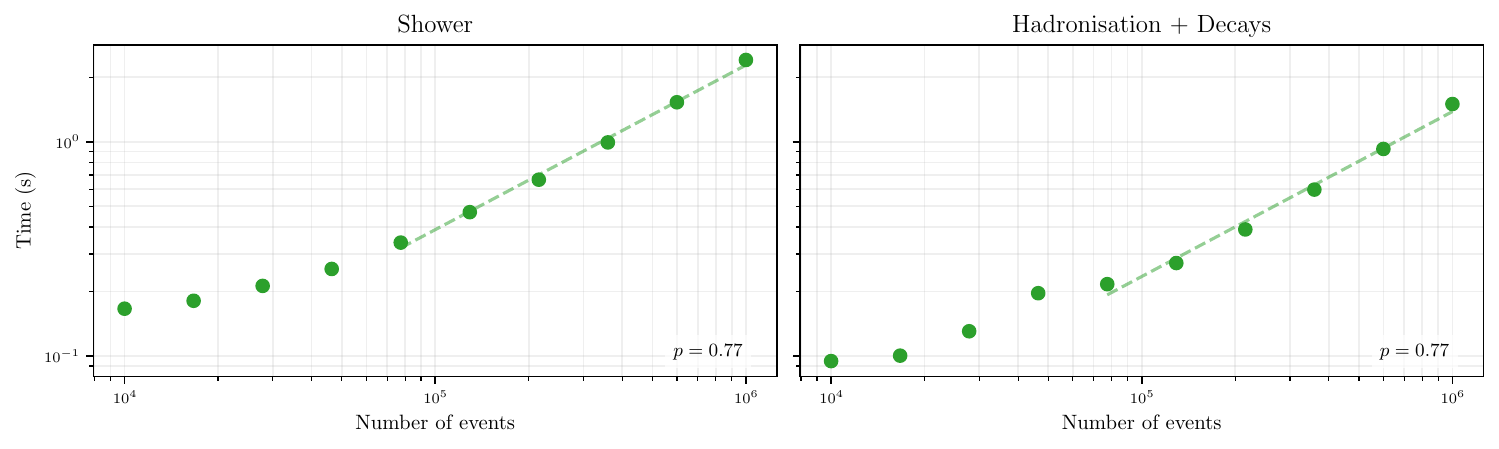}
\caption{Execution times on an NVIDIA V100 GPU for the shower and hadronisation+decays. Points are medians of ten runs; the interquartile ranges are smaller than the markers. The dashed lines are power-law fits to $N \geq 5 \times 10^4$ events. Runs on more powerful GPUs or multi-GPU systems would show whether the power-law trend continues beyond $10^6$ events.}
\label{fig:component-times}
\end{figure*}

\section{\label{sec:results}Hadron-Level Results}
I simulated $10^6$ events of the hard process $e^+e^- \to q \bar{q}$ at NLO for $\sqrt{s} = 91.2\,\text{GeV}$, with showering and hadronisation, and with hadronic decays on and off.
For physics benchmarking, I ran the same simulations with Herwig 7.3 \cite{Bewick:2023tfi}.
Both GAPS and Herwig match the NLO calculation to the shower with MC@NLO matching \cite{Frixione:2002ik, Platzer:2011bc}.
In both codes, I used the CMW scheme \cite{Catani:1990rr} for the shower $\alpha_{\text{s}}$, with $\alpha_{\text{s}}(m_Z) = 0.118$, and a shower cutoff of $p_{T,\min}^2 = 1\,\text{GeV}^2$.
Both codes use the hadronisation parameters in Table~\ref{tab:had-params}: the cluster parameters and flavour weights are Herwig's default values, not retuned for this setup.
In Herwig, I switched off colour reconnection, used the older fission algorithm and isotropic cluster decays, and set the singlet and decuplet suppression factors to 1, to match GAPS.

\begin{table}
\centering
\begin{tabular}{l|ccc}
\hline
 & Light & Charm & Bottom \\
\hline
\texttt{Clmax} (GeV) & 3.53 & 3.95 & 3.76 \\
\texttt{Clpow} & 1.85 & 2.56 & 0.55 \\
\texttt{PSplit} & 0.91 & 0.99 & 0.63 \\
\hline
\end{tabular}

\vspace{0.5em}
\begin{tabular}{l|ccccccc}
\hline
 & $d$ & $u$ & $s$ & $c$ & $b$ & $g$ & $qq$ \\
\hline
$m$ (GeV) & 0.33 & 0.33 & 0.45 & 1.30 & 4.75 & 0.95 & -- \\
$w$ & 1.0 & 1.0 & 0.37 & -- & -- & -- & 0.33 \\
\hline
\end{tabular}
\caption{Hadronisation parameters used in both GAPS and Herwig: the fission parameters for light, charm and bottom clusters, and the constituent masses $m$ and flavour weights $w$ of the quarks, the gluon and the diquark.}
\label{tab:had-params}
\end{table}

I used the final states to compute event shapes, jet rates and multiplicities.
I first compare the two generators without hadronic decays and then with decays, where I also compare them with ALEPH and L3 data \cite{ALEPH:2003obs, L3:2004cdh}.
Figure~\ref{fig:hadron-level} presents these results, with plots made using Rivet \cite{Bierlich:2019rhm}.
GAPS computes the observables with its own implementation of the ALEPH and L3 analyses, which I validated against Rivet's.
GAPS and Herwig agree with each other and with the data for event shapes and jet rates, including their hard-$p_T$ tails, with the Catani-Seymour dipole subtraction \cite{Catani:1996vz} and shower \cite{Schumann:2007mg, Platzer:2009jq} providing the main contribution to this agreement.
The exception is the two-jet limit, where GAPS lies below Herwig.
For the charged multiplicity, GAPS and Herwig agree well for $8 \lesssim N_{\text{ch}} \lesssim 30$ and deviate above this range.
Before hadronic decays, the two codes give similar overall multiplicities, so this deviation most likely stems from the simplified hadronic decay model.
Events that overflow the GAPS event record have the highest multiplicities and are excluded from the histograms, so excluding them further depletes the tail.

\section{\label{sec:times}Execution Times and Profiling}

As in previous releases, I measured the code's computational performance on an NVIDIA V100 GPU \cite{nvidia-tesla-2024}.
First, to see how each component contributes to the total event generation time, I profiled the simulation of $10^6$ events, with the timings presented in Table~\ref{tab:cuda_kernels}.
Hadronic decays and cluster decay are the most time-consuming of the hadronisation and decay kernels.
However, they consume less time than observable computations: increased multiplicity significantly slows the thrust calculation and jet clustering, motivating better-scaling algorithms, such as those in FastJet \cite{Cacciari:2005hq, Cacciari:2011ma} for jet clustering.
That said, the observable computations are auxiliary to event generation, and improving their performance is beyond the scope of this note.

Next, I measured how execution time scales with the number of events, separately for the shower and hadronisation+decays stages, as shown in Figure~\ref{fig:component-times}.
The shower time grows as a power law $t \propto N^p$ in the number of events $N$, with $p = 0.77$ above about $5 \times 10^4$ events, the same sublinear scaling seen in previous releases \cite{Seymour:2024fmq, Seymour:2025fpu}.
The same holds for the hadronisation and decay algorithms, which take time comparable to the shower, so adding them does not change the overall GPU simulation scaling.
At $10^6$ events, the matrix element, shower, hadronisation and decays take about four seconds in total, excluding device preparation.

\begin{table}
\centering
\begin{tabular}{l|r|r|r}
\hline
Name & Calls & Time (ms) & Time (\%) \\
\hline
Event Shapes & 1 & 6241.3 & 36.2 \\
Durham Jets & 1 & 4270.9 & 24.8 \\
\textit{Device Prep} & 1 & 2664.4 & 15.5 \\
Trial Emission & 158 & 2102.9 & 12.2 \\
Hadronic Decays & 1 & 1154.7 & 6.7 \\
Fill Histograms & 1 & 201.2 & 1.2 \\
Cluster Decay & 1 & 195.3 & 1.1 \\
Perform Emission & 158 & 89.8 & 0.5 \\
Cluster Fission & 1 & 55.6 & 0.3 \\
Const. Reshuffling & 1 & 55.1 & 0.3 \\
\hline
\end{tabular}
\caption{The ten largest contributions to the GPU run time for $10^6$ events, profiled using Nsight Systems \cite{nvidia-nsight-2025}.
The time consumed by the shower exceeds that reported for the first release of GAPS \cite{Seymour:2024fmq}, which used a smaller event record and a final-state-only shower at LO.
\textit{Device Prep} is the sum of the one-off kernels that initialise the event records, which take most of this time, and the matrix-element, shower and hadronisation objects; note that this cost is small compared to showering LHC events, which take around a minute \cite{Seymour:2025fpu}.}
\label{tab:cuda_kernels}
\end{table}

\section{\label{sec:conc}Conclusions and Outlook}

This note presents a GPU implementation of cluster hadronisation and hadronic decays in GAPS.
Before hadronic decays, the hadron-level event shapes and jet rates agree closely with Herwig, except in the two-jet limit, where GAPS lies $15\%$ below Herwig.
After decays, the charged multiplicity agrees well with Herwig for $8 \lesssim N_{\text{ch}} \lesssim 30$, and the remaining differences most likely come from the simplified hadronic decay model and the exclusion of events that overflow the event record.
Computationally, hadronisation on the GPU scales with a similar power law to the shower, so it does not change the scaling of the GPU simulation.
The code is available as release 2.2 of GAPS,
\begin{center}
\texttt{\url{https://gitlab.com/siddharthsule/gaps}}
\end{center}

Several steps remain before GAPS is a complete GPU event generator.
The decay model omits partonic decays of heavy hadrons and uses flat phase space; a complete decay model using decay matrix elements will be the goal of a future release.
The hadronisation and decay kernels also lack GPU-specific optimisation, such as reducing thread divergence in the fission and decay loops.
Additionally, hadronisation of $pp$ and DIS events requires the modelling of beam remnants.
I also expect the multiplicity to increase further, requiring a higher $N_{\max}$, which may reduce the number of events that can be simulated in parallel on the GPU.
Events that overflow the event record will also need proper handling rather than exclusion.

\begin{acknowledgments}
I developed the GAPS codebase in collaboration with M. H. Seymour.
I thank him, together with A. Si\'odmok and J. Whitehead, for useful discussions and for comments on this note. 
I acknowledge support from the UK Science and Technology Facilities Council (STFC) through a studentship.
I thank the University of Manchester for access to the Noether Cluster, which was used to generate the results in this note.

\textit{AI Tools Declaration:} The physics implemented in the generator reflects my understanding of the Herwig cluster algorithm, and I designed any modifications myself. I used Claude Code to help debug the code and refine code comments.
\end{acknowledgments}

\makeatletter
\def\bibsection{%
  \par
  \begingroup\baselineskip26\p@\bib@device{\hsize}{72\p@}\endgroup
  \nobreak\@nobreaktrue
  \addvspace{19\p@}%
}
\makeatother

\bibliographystyle{jhep-customised}
\bibliography{main}

\end{document}